\documentclass[conference]{IEEEtran}
\usepackage{amsfonts}
\usepackage{mathrsfs}
\usepackage{amssymb,amsmath}
\usepackage{stfloats}
\usepackage{algorithm}
\usepackage{algorithmic}
\usepackage{amsmath,amssymb,epsfig,graphics,subfigure}
\usepackage{theorem}
\usepackage{array,color}
\usepackage[compress]{cite}
\usepackage{url}
\usepackage{slashbox,multirow}

\newcommand{\name}{{\tt VAE-CIR}}
\newcommand{\nameS}{{\tt VAE-CIR} }

\theoremheaderfont{\normalfont\bfseries}

\begin{document}
\title{Accurate Angular Inference for 802.11ad Devices Using Beam-Specific Measurements}

\author{ \IEEEauthorblockN{Haichuan Ding and  Kang G. Shin}
\IEEEauthorblockA{University of Michigan, Ann Arbor \\ Email:dhcbit@gmail.com, kgshin@umich.edu}}
\maketitle

\begin{abstract}
Due to their sparsity, 60GHz channels are characterized by a few dominant paths. Knowing the angular information 
of their dominant paths, we can develop various applications, such as the prediction of link performance and 
the tracking of an 802.11ad device. Although they are equipped with phased arrays, the angular
inference for 802.11ad devices is still challenging due to their limited number of RF chains and limited
phase control capabilities. 
Considering the beam sweeping operation and the high communication bandwidth of 802.11ad devices, 
we propose variation-based angle estimation (VAE), called \name,  by utilizing beam-specific channel impulse 
responses (CIRs) measured under different beams and the directional gains of the corresponding beams to 
infer the angular information 
of dominant paths. Unlike state-of-the-arts, \nameS exploits the variations between different beam-specific CIRs, 
instead of their absolute values, for angular inference. To evaluate the performance of \name, we generate the 
beam-specific CIRs by simulating the beam sweeping of 802.11ad devices with the beam patterns measured 
on off-the-shelf 802.11ad devices. The 60GHz channel is generated via a ray-tracing simulator 
and the CIRs are extracted via channel estimation based on Golay sequences. Through experiments in 
various scenarios, we demonstrate the effectiveness of \nameS and its superiority to existing angular 
inference schemes for 802.11ad devices.
\end{abstract}



\section{Introduction}
The completion of the 802.11ad standard and its device commercialization have led to various research efforts 
to either improve the communication performance of 60GHz devices, such as link performance prediction, 
access point (AP) deployment, and seamless indoor multi-Gbps wireless connectivity, or develop novel 
applications with these devices, such as single AP decimeter-level localization 
\cite{wei2017facilitating,sur2018towards,zhou2017beam,ghasempour2018multi,pefkianakis2018accurate}. 
All these applications usually rely on the length and angular information of strong signal (or {\em dominant})
paths in the sparse 60GHz channel. Since this path information is often unavailable, we must be able to 
effectively infer such information for which angular information serves as the starting point.

As demonstrated in various studies, the starting point for inferring path information is the collection of angular 
information --- such as angles of departure (AODs) and angles of arrival (AOAs) --- of dominant paths 
\cite{wei2017facilitating,sur2018towards,zhou2017beam,ghasempour2018multi,pefkianakis2018accurate}. 
Unfortunately, inferring angular information is not easy for 802.11ad devices. On the one hand, for cost reasons, 
phased arrays on off-the-shelf 802.11ad devices only support 2 to 4-bit phase control \cite{sur2017wifi}. 
Due to this limited phase control capability, the generated beams can have multiple strong side lobes, thus making 
it difficult for 802.11ad devices to directly learn angular information from the adopted beams. 
On the other hand, due to the limited number of RF chains, existing direction estimation schemes are not applicable 
to off-the-shelf 802.11ad devices. As reported in \cite{wei2017facilitating}, 802.11ad devices adopt analog phased 
arrays and the signals received at different antenna elements are mixed at the output of the RF chain. 
Since most existing direction estimation schemes require the signal received at each antenna element to be known, 
they are not applicable to millimeter wave (mmWave) devices with analog phased arrays. Although there have been a few direction 
estimation schemes proposed for antenna arrays with a hybrid beamforming architecture, they cannot be directly 
integrated into 802.11ad devices which are equipped with a single RF chain. 

Considering this fact, several direction estimation schemes have recently been proposed for 802.11ad devices 
based on a single RF chain and analog arrays. Wei {\em et al.}~\cite{wei2017facilitating} exploited the orthogonality 
within the codebook to map the output of the RF chain to the signals received at different antenna elements and 
used the MUSIC algorithm to estimate the AOAs of different paths. In their scheme, the device receives the 
incoming signal using different weight vectors (i.e., beam patterns) and isolates the signal received at each antenna 
element through a matrix inversion. As pointed out in \cite{pefkianakis2018accurate}, to make this scheme effective, 
we should ensure that exactly the same signal is sampled by all of the adopted beam patterns, thus limiting its applicability. Furthermore, the placement and enclosures of the phased array antenna might affect its radiation 
pattern, making the relationship between weight vectors and directional gains invalid, and thus adversely impacting the 
performance of the  AOA estimation schemes based on weight vectors \cite{hosoya2015multiple}. So, most existing 
direction estimation schemes for 802.11ad devices exploit the correlation between wireless measurements 
made under different beams and the directional gains of the corresponding beams for angular inference 
\cite{steinmetzer2017compressive,pefkianakis2018accurate,ghasempour2018multi}.  
Steinmetzer {\em et al.}~\cite{steinmetzer2017compressive} proposed a direction estimation scheme based on
received signal strength (RSS) to infer the AOA of a dominant path. In that scheme, multiple RSS measurements 
are made using different beams, and the estimated AOA is the direction that provides the best match between the 
directional gains and the RSS measurements. Although its effectiveness has been demonstrated in a specific 
environment, the experimental results in \cite{pefkianakis2018accurate} show this RSS-based scheme to be highly 
inaccurate in a different environment due to irregular beam patterns and multipath propagation. In this case, 
the RSS not only depends on the path of interest but also is affected by the signals received from other paths.

This observation motivates two recent direction estimation schemes which exploit the measured channel impulse 
response (CIR), instead of the RSS, for angular inference \cite{ghasempour2018multi,pefkianakis2018accurate}. 
According to \cite{IEEE2012adIEEE}, each 802.11ad frame begins with a short training field (STF) and a channel 
estimation (CE) field, which allow 802.11ad devices to acquire the corresponding CIR. Additionally, since the 
802.11ad frames are delivered over a spectrum of  $\geq 1760 MHz$ bandwidth, two path components are resolvable 
for 802.11ad devices as long as their arrival times have a $0.57ns$ difference. In other words, two paths are 
resolvable as long as their lengths have a $0.17m$ difference. Given the sparsity of mmWave channels, it is very
likely that each dominant component in the CIR corresponds to one of the dominant propagation paths between the 
transmitter and the receiver. Thus, each dominant CIR component is very likely to be completely determined by a 
dominant path. In view of this, Ghasempour {\em et al.}~\cite{ghasempour2018multi} and Pefkianakis 
{\em et al.}~\cite{pefkianakis2018accurate} propose two direction estimation schemes for 802.11ad devices, 
which exploits beam-specific CIRs measured under different beam patterns during, for example, beam 
sweeping to gradually narrow down the potential AOAs/AODs. 
In both of these two schemes, CIR components are first mapped to dominant paths through their time 
of arrival. Then, the corresponding CIR components measured under different beams are used to gradually refine the 
AOA/AOD estimates of the paths of interest. For each dominant path, the scheme in \cite{ghasempour2018multi} 
maintains a score for each direction with initial value $0$. Once a new CIR measurement comes in, the score of each 
direction is updated depending on whether the desired CIR component is received or not, as well as the directional gain 
of the adopted beam pattern. When the measurement process completes, the direction with the highest score will be 
considered as the AOA of the corresponding path. The authors of \cite{pefkianakis2018accurate} adopted a similar 
procedure to estimate the AOD of the line-of-sight (LOS) path by utilizing the CIR feedback from the receiver. 
To improve the accuracy of AOD estimation further, they selected a few CIR measurements with the strongest LOS 
components for AOD estimation and weighted the contribution of each CIR measurement with the amplitude of its LOS 
component. They then chose the estimated AOA to be the center of the direction interval spanned by a few directions 
with the highest scores. Despite the differences, both of these schemes rely on a hidden assumption that, 
if a CIR component is received under a beam, the angle of the corresponding path is more likely to be along the 
direction with a higher directional gain. As a result, once the desired CIR component is measured under a beam, 
they always assign the highest score increase to the direction with the highest directional gain. 
However, the reception of a CIR component only implies that the directional gain along the AOA/AOD of the 
corresponding path is higher than a certain threshold, rather than that the AOA/AOD of this path is along the 
direction with the highest gain. Blindly assigning higher score increments to the directions with higher 
directional gains might lead to large estimation errors.

As seen from the above discussions, despite the various angle estimation schemes proposed thus far, accurate 
AOA/AOD estimation for 802.11ad devices is still lacking. The goal of this paper is to address this need by 
developing a variation-based angle estimation (VAE), called \name. 
Like \cite{pefkianakis2018accurate,ghasempour2018multi}, for each path of interest,  \nameS maintains 
the score for each direction and utilizes the beam-specific CIR measurements under different beam patterns 
to update the score. Unlike existing schemes where the CIR measured under each beam is used independently 
to update the score, \nameS updates the score based on the variations 
in the corresponding CIR component under different beams. Specifically, for the considered path, \nameS updates 
the score for each direction based on the variations between every two CIRs measured using different beams, 
and the direction where the variation in directional gain better matches that in the CIR will be
assigned a higher score increase.
In other words, \nameS picks the direction in which the variations of directional gain best match the 
variations in the CIR component as the estimated angle for the path of interest. The rationale behind \nameS is that, 
during measurement processes like beam sweeping,
different beams encounter the same set of signal paths and the variations in CIR measurements are the results of 
the variations in directional gain along the AOA/AOD of each path. This paper makes the following
contributions.

\begin{itemize}
\item Study  of the angular inference for 802.11ad devices based on wireless measurements. Unlike prior work, 
we propose to not directly utilize each CIR component but employ its variations under different beams 
to infer the angle of each path.

\item Design of \nameS which exploits the variations in every two CIR measurements to infer the AOA/AOD 
of dominant paths. For a specific path, \nameS exploits the variations in its corresponding CIR component 
between every two measurements to update the score assigned to different directions and the inferred AOA/AOD 
is the direction with the highest final score. \nameS allows us to flexibly weight the contribution of every two CIR 
measurements to the score. For example, a lower weight can be assigned when the amplitude of the interested 
CIR component is below a certain threshold in at least one of the two CIR measurements since the accuracy 
of the CIR measurements is more prone to noise in this case.
 
\item Extensive simulation to evaluate effectiveness of \name. We generate the transmit signals and implement 
the channel estimation based on Golay sequences for CIR extraction by following the IEEE 802.11ad standard. 
Using the beam pattern measured on the off-the-shelf 802.11ad devices and the channel model approved 
by IEEE 802.11 Task Group ad, we demonstrate the superiority of \nameS to state-of-the-arts.
\end{itemize}

\section{Beam-Specific CIR and Channel Estimation}
\subsection{Beam-Specific CIR}
To compensate high path loss and facilitate high speed transmissions, mmWave communication devices exploit 
phased arrays and beamforming to concentrate energy in the desired directions. Due to the constraints on hardware 
implementation, 802.11ad devices adopt an analog beamforming architecture where each antenna is connected with 
a phase shifter and beamforming is performed in the analog domain with these phase shifters. This analog 
beamforming architecture cannot simultaneously support more than one beam, and can steer the beam to 
different directions by controlling the amount of phase shift of each phase shifter \cite{ahmed2018survey}. 
Due to the use of beamforming, the CIR observed at the receiver depends on both the underlying signal paths 
and the radiation pattern of the adopted beam. Denoting the equivalent lowpass representation of the 
continuous-time CIR as $h(t)$ \cite{goldsmith2005wireless}, we have
\begin{align}
\label{1}
h\left( t \right) = \sum\limits_{k = 1}^K {\alpha _k \mathcal{G}_m \left( {\vartheta _k } \right)G_l \left( {\theta _k } \right)\delta \left( {t - \tau _k } \right)} ,
\end{align} 
where $K$ is the total number of underlying signal paths, $\alpha _k$ and and $\tau_k$ are the complex path gain
and the delay of the $k$-th path,  respectively. $\vartheta _k$ is the AOD of the $k$-th path at the transmitter and 
$\mathcal{G}_m \left( {\vartheta _k } \right)$ is the beamforming gain along the spatial direction $\vartheta _k$ when the 
transmitter uses its $m$-th beam for transmissions. $\theta _k$ is the AOA of the $k$-th path at the receiver and 
$G_l \left( {\theta _k } \right)$ is the beamforming gain along direction $\theta _k$ when the receiver employs its $l$-th 
beam for reception. Each summand in Eq.~(\ref{1}) corresponds to a CIR component. Let $h _k$ be the $k$-th 
CIR component and $h _k\left(l\right)$ be the $k$-th CIR component measured using the $l$-th receiving beam. 
When the underlying signal path remains the same, the amplitude of each CIR component, $\left|h _k\right|$, will 
track the changes of directional gains. For example, if the receiver switches to the $(l+1)$-th beam, we have 
$\left| h _k \right|=\left| h _k \left( l+1 \right) \right|=\left| {\alpha _k \mathcal{G}_m \left( {\vartheta _k } \right)G_{l + 1}
 \left( {\theta _k } \right)} \right|$ and ${{\left| h _k \left( {l + 1} \right) \right|} \mathord{\left/
 {\vphantom {{ \left| h _k \left( {l + 1} \right) \right|} {\left|h _k \left( l \right)\right|}}} \right.
 \kern-\nulldelimiterspace} {\left|h _k \left( l \right) \right|}} = {{\left| {G_{l + 1} \left( {\theta _k } \right)} \right|} \mathord{\left/
 {\vphantom {{\left| {G_{l + 1} \left( {\theta _k } \right)} \right|} {\left| {G_l \left( {\theta _k } \right)} \right|}}} \right.
 \kern-\nulldelimiterspace} {\left| {G_l \left( {\theta _k } \right)} \right|}}$. In other words, the variations in the directional 
 gain along the AOA $\theta _k$ of a path matches the variations in the amplitude of its corresponding CIR 
 component $h_k$. This observation motivates \nameS which we introduce in the next section.

\subsection{Channel Estimation in 802.11ad}

\begin{figure}[!t]
\begin{center}
  \includegraphics[width=3.5in]{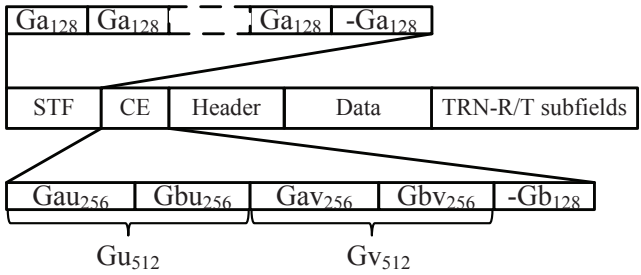}
  \end{center}
  \begin{center}
   \parbox{8cm}{\caption{The structure of a typical 802.11ad PPDU.}  \label{fig}}
  \end{center}
\end{figure}

According to the IEEE 802.11ad standard, each physical layer convergence procedure protocol data unit (PPDU) 
contains a CE field right next to the STF. 802.11ad physical layer (PHY) supports three modulation methods --- 
control modulation, single carrier (SC) modulation and OFDM modulation --- which share a common preamble 
structure \cite{IEEE2012adIEEE}. The structure of a typical 802.11ad PPDU is shown in Fig.~\ref{fig}, where 
$Gau_{256}[n]$, $Gbu_{256}[n]$, $Gav_{256}[n]$, $Gbv_{256}[n]$ are Golay sequences of length $256$. 
$Ga_{128}(n)$ and $Gb_{128}(n)$ are two Golay sequences of length $128$ defined in subclause 21.11 of the 
802.11ad standard. $Gau_{256}(n)$ and $Gbu_{256}(n)$ are complementary sequences with the following 
property:
\begin{align}
\label{2}
R_a \left[ i \right] + R_b \left[ i \right] = \left\{ {\begin{array}{*{20}c}
   {512} & {i = 0}  \\
   0 & { - 255 \le i \le 255,i \ne 0}  \\
\end{array}} \right.,
\end{align}
where $R_a \left[ i \right]$ and $R_b \left[ i \right]$ are autocorrelation of $Gau_{256}\left[n\right]$ and $Gbu_{256}\left[n\right]$, defined as
\begin{align}
\label{3}
 R_a \left[ i \right] =&\left(Gau_{256} \star Gau_{256}\right)\left[i\right] \nonumber \\=& \sum\limits_{n = \max \left\{ {0,i} \right\}}^{\min \left\{ {255 + i,255} \right\}} {Gau_{256} \left[ n \right]Gau_{256} \left[ {n - i} \right]}, \nonumber \\ &-255 \le i \le 255, \nonumber \\ 
 R_b \left[ i \right] =& \left(Gbu_{256} \star Gbu_{256}\right)\left[i\right] \nonumber \\=& \sum\limits_{n = \max \left\{ {0,i} \right\}}^{\min \left\{ {255 + i,255} \right\}} {Gbu_{256} \left[ n \right]Gbu_{256} \left[ {n - i} \right]}, \nonumber \\ &-255 \le i \le 255,
\end{align}
where $\star$ is the correlation operator. Similarly, $Gav_{256}(n)$ and $Gbv_{256}(n)$ are complementary 
sequences. $Ga_{128}(n)$ and $Gb_{128}(n)$ are complementary sequences. These two pairs of complementary 
sequences also have the autocorrelation property in Eq.~(\ref{2}), and this property of Golay sequences is exploited 
in 802.11ad for channel estimation \cite{mishra2017sub,liu2013digital}. Clearly from Fig.~\ref{fig}, both the STF 
and CE field are built up from Golay sequences. Each STF ends with $-Ga_{128}(n)$, and each CE field ends 
with $-Gb_{128}(n)$, which facilitates the creation of a length-127 zero-correlation zones as discussed below 
\cite{yong201160ghz}. It should be noted that the CE field shown in Fig.~\ref{fig} is the CE field for control PHY 
and SC PHY. OFDM PHY adopts a similar CE field where $Gv_{512}[n]$ is transmitted before $Gu_{512}[n]$. 
The STF and CE fields are modulated prior to transmission using 
${\pi  \mathord{\left/ {\vphantom {\pi  2}} \right. \kern-\nulldelimiterspace} 2}$-BPSK. 
We use $\widetilde{Gau}_{256} \left[ n \right]$ and $\widetilde{Gbu}_{256} \left[ n \right]$ to denote 
${Gau_{256} \left[ n \right]}$ and ${Gbu_{256} \left[ n \right]}$ after constellation mapping. 
It can be easily shown that $\widetilde{Gau}_{256} \left[ n \right]$ and $\widetilde{Gbu}_{256} \left[ n \right]$ 
also have the autocorrelation property shown in Eq.~(\ref{2}).

Let $s(t)$ be the transmitted preamble before upconversion. Then, we have 
\begin{align}
\label{4}
s\left( t \right) = \sum\limits_{n = 0}^{1279} {s\left[ n \right]\left( {g * h_T } \right)\left( {t - nT_c } \right)},
\end{align}
where $s\left[ n \right]$ is a sequence obtained by sequentially concatenating $-{\widetilde{Ga}_{128} }$, 
${\widetilde{Gu}_{512} }$, ${\widetilde{Gv}_{512} }$, and $-{\widetilde{Gb}_{128} }$, $h_T\left( t \right)$ represents the transmit shaping filter, 
and $T_c$ is the chip time. $T_c=0.57ns$ for control and SC PHY. 
$g\left( t \right)=u\left( t \right) - u\left( {t - T_c } \right)$, where $u\left( t \right)$ is the unit step function defined 
in \cite{oppenheim1997signals}. $*$ in Eq.~(\ref{4}) represents the operation of convolution. It should be noted 
that, for ease of presentation, we only consider the part of the preamble related to channel estimation in 
Eq.~(\ref{4}). Then, the received signal after going through the wireless channel, downcoversion, and 
the anti-aliasing filter is
\begin{align}
\label{5}
r\left( t \right) = \sum\limits_{k = 1}^K {h_k \left( {s * h_R } \right)\left( {t - \tau _k } \right)}  + z\left( t \right),
\end{align}
where $h_k$ is the $k$th CIR component defined in the last subsection, ${h_R }\left( t \right)$ is the impulse 
response of the anti-aliasing filter at the receiver, and $z\left( t \right)$ is the additive noise. Suppose 
$\left( {h_T  * h_R } \right)\left( t \right)$ has a flat frequency response with magnitude $1$ in the band of interest, 
the received signal sampled at a rate of $
{1 \mathord{\left/ {\vphantom {1 {T_c }}} \right.
 \kern-\nulldelimiterspace} {T_c }}$ is
\begin{align}
\label{6}
r\left( {nT_c } \right) = \sum\limits_{k = 1}^K {h_k \sum\limits_{j = 0}^{1279} {s\left[ j \right]g\left( {nT_c  - \tau _k  - jT_c } \right)} }  + z\left( {nT_c } \right).
\end{align}
Let $r\left[ n \right] = r\left( {nT_c } \right)$ and $z\left[ n \right] = z\left( {nT_c } \right)$, we have
\begin{align}
\label{7}
r\left[ n \right] = \sum\limits_{k = 1}^K {h_k s\left[ {n - \left\lceil {{{\tau _k } \mathord{\left/
 {\vphantom {{\tau _k } {T_c }}} \right.
 \kern-\nulldelimiterspace} {T_c }}} \right\rceil } \right]}  + z\left[ n \right],
\end{align}
where $\left\lceil . \right\rceil$ is the ceiling function. To obtain the estimated CIR, $r\left[ n \right]$ is first correlated 
with ${\widetilde{Gau}_{256} }$, ${\widetilde{Gbu}_{256} }$, ${\widetilde{Gav}_{256} }$, and 
${\widetilde{Gbv}_{256} }$. Specifically, we have
\begin{align}
\label{8}
\left( {r \star \widetilde{Gau}_{256} } \right)\left[ i \right] = & \sum\limits_{k = 1}^K {h_k \left( {s \star \widetilde{Gau}_{256} } \right)\left[ {i - \left\lceil {{{\tau _k } \mathord{\left/
 {\vphantom {{\tau _k } {T_c }}} \right.
 \kern-\nulldelimiterspace} {T_c }}} \right\rceil } \right]}  \nonumber \\ &+ \left( {z \star \widetilde{Gau}_{256} } \right)\left[ i \right], 
\end{align}
where
\begin{align}
\label{9}
&\left( {s \star \widetilde{Gau}_{256} } \right)\left[ i \right] \nonumber \\ &= \left\{ {\begin{array}{*{20}c}
   {\sum\limits_{n = \max \left\{ { - i,0} \right\}}^{\min \left\{ {\ell _s  - i,255} \right\}} {s\left[ n \right]\widetilde{Gau}_{256} \left[ {n - i} \right]} } & { - 255 \le i \le \ell _s }  \\
   0 & {o.w.}  \\
\end{array}} \right..
\end{align}
Similar results can be obtained for ${\widetilde{Gbu}_{256} }$, ${\widetilde{Gav}_{256} }$, and 
${\widetilde{Gbv}_{256} }$.
 
\begin{figure}[!t]
\begin{center}
  \includegraphics[width=3.5in]{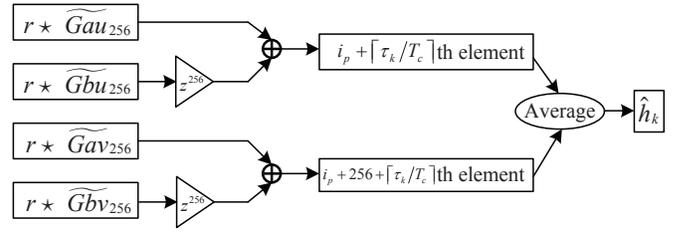}
  \end{center}
  \begin{center}
   \parbox{8cm}{\caption{Golay sequence based estimation for the $k$-th CIR component. 
   $z^{256}$ is the delay operator.}  \label{fig1}}
  \end{center}
\end{figure}

\begin{figure}[!t]
\begin{center}
  \includegraphics[width=3in]{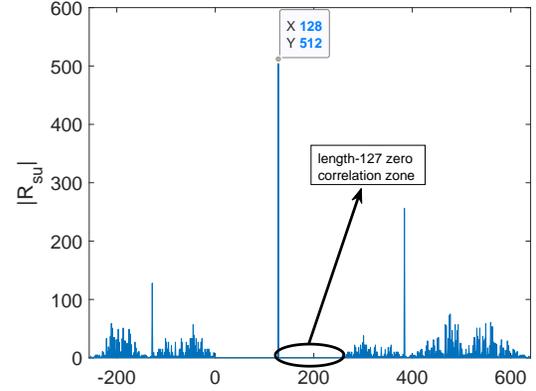}
  \end{center}
  \begin{center}
   \parbox{8cm}{\caption{The value of $\left| {R_{su} \left[ i \right]} \right|$. We have $i_p=128$ since $\widetilde{Gau}_{256}$ starts at the $128$th element of $s$. A length-127 zero correlation zone can be clearly seen before and after the index of $128$.}  \label{fig2}}
  \end{center}
\end{figure}

As shown in Fig.~\ref{fig1}, $\left( {r \star \widetilde{Gau}_{256} } \right)\left[ i \right]$ is then combined with 
$\left( {r \star \widetilde{Gbu}_{256} } \right)\left[ i \right]$, $\left( {r \star \widetilde{Gav}_{256} } \right)\left[ i \right]$, 
and $\left( {r \star \widetilde{Gbv}_{256} } \right)\left[ i \right]$ to obtain the estimated CIR. 
By adding $\left( {r \star \widetilde{Gau}_{256} } \right)\left[ i \right]$ and $\left( {r \star \widetilde{Gbu}_{256} } 
\right)\left[ i+256 \right]$, we have
\begin{align}
\label{10}
&R_{ru} \left[ i \right] = \sum\limits_{k = 1}^K {h_k R_{su}\left[ {i - \left\lceil {{{\tau _k } \mathord{\left/
 {\vphantom {{\tau _k } {T_c }}} \right.
 \kern-\nulldelimiterspace} {T_c }}} \right\rceil } \right]}  + \widetilde{z}\left[ i \right],
\end{align}
where $\widetilde{z}\left[ i \right]$ is the noise after correlation operation and
\begin{align}
\label{11}
R_{ru} \left[ i \right]=\left( {r \star \widetilde{Gau}_{256} } \right)\left[ i \right]  +\left( { r \star \widetilde{Gbu}_{256} } \right)\left[ i+256 \right], \nonumber \\
{R_{su} \left[ i \right]}=\left( {s \star \widetilde{Gau}_{256}} \right)\left[ i \right]  + \left( {s \star \widetilde{Gbu}_{256} } \right)\left[ i +256\right].
\end{align}
Clearly from Eq.~(\ref{10}), the value of $R_{ru} \left[ i \right]$ is closely related to that of ${R_{su} \left[ i \right]}$. Let $R_{sau} \left[ i \right]=\left( {s \star 
\widetilde{Gau}_{256} } \right)\left[ i \right]$, and $R_{sbu} \left[ i \right]=\left( {s \star \widetilde{Gbu}_{256} } 
\right)\left[ i \right]$. By exploiting the property of Golay sequences shown in Eq.~(\ref{2}), we have 
\begin{align}
\label{12}
\left| {R_{su} \left[ i \right]} \right|=&\left|R_{sau} \left[ i \right] + R_{sbu} \left[ {i + 256} \right] \right| \nonumber \\ =
& \left\{ {\begin{array}{*{20}c}
   {512} & {i = 128}  \\
   0 & {0 \le i \le 255,i \ne 128}  \\
\end{array}} \right.,
\end{align} 
From Eq.~(\ref{12}), $\left| {R_{su} \left[ i \right]} \right|$ has a peak at the index $i_p=128$ where 
$\widetilde{Gau}_{256}$ aligns with $\widetilde{Gau}_{256}$ in $s$, and there is a length-127 zero-correlation 
zone before and after the peak. This matches the result in Fig.~\ref{fig2} well. 

Noting that $R_{su}\left[ {i - \left\lceil {{{\tau _k } \mathord{\left/
 {\vphantom {{\tau _k } {T_c }}} \right.
 \kern-\nulldelimiterspace} {T_c }}} \right\rceil } \right]$ is a shifted version of $R_{su}\left[i\right]$, $\left|R_{su}
 \left[ {i - \left\lceil {{{\tau _k } \mathord{\left/ {\vphantom {{\tau _k } {T_c }}} \right.
 \kern-\nulldelimiterspace} {T_c }}} \right\rceil } \right]\right|$ has a peak at location $i_p\left( {k} \right)= i_p + {\left\lceil {{{\tau _k } \mathord{\left/
 {\vphantom {{\tau _k } {T_c }}} \right.
 \kern-\nulldelimiterspace} {T_c }}} \right\rceil }$. Without of loss generality, let us assume $\left\lceil {{{\tau _{k'} } \mathord{\left/
 {\vphantom {{\tau _{k'} } {T_c }}} \right.
 \kern-\nulldelimiterspace} {T_c }}} \right\rceil  \ne \left\lceil {{{\tau _k } \mathord{\left/
 {\vphantom {{\tau _k } {T_c }}} \right.
 \kern-\nulldelimiterspace} {T_c }}} \right\rceil $ if $k' \ne k$ and $\left| {\tau _K  - \tau _1 } \right| \le 128T_c$. 
 Then, if the noise is neglected, we get $\left| {R_{ru} \left[ {i_p  + \left\lceil {{{\tau _k } \mathord{\left/
 {\vphantom {{\tau _k } {T_c }}} \right.
 \kern-\nulldelimiterspace} {T_c }}} \right\rceil } \right]} \right|=512 \left|h_k\right|$. 
 In other words, we can recover the $k$-th CIR component by extracting the ${i_p  + \left\lceil {{{\tau _k } \mathord{\left/
 {\vphantom {{\tau _k } {T_c }}} \right.
 \kern-\nulldelimiterspace} {T_c }}} \right\rceil }$th element in $R_{ru}\left[ i \right]$ through $\widehat{h}_k^u  = {{R_{ru} \left[{i_p  + \left\lceil {{{\tau _k } \mathord{\left/
 {\vphantom {{\tau _k } {T_c }}} \right.
 \kern-\nulldelimiterspace} {T_c }}} \right\rceil } \right]} \mathord{\left/
 {\vphantom {{R_{ru} \left[{i_p  + \left\lceil {{{\tau _k } \mathord{\left/
 {\vphantom {{\tau _k } {T_c }}} \right.
 \kern-\nulldelimiterspace} {T_c }}} \right\rceil }\right]} {512}}} \right.
 \kern-\nulldelimiterspace} {512}}$, where $\widehat{h}_k^u $ represents the estimate of the $k$-th CIR component. Following the same procedure, we can obtain another estimate of the $k$-th CIR component, denoted as 
 $\widehat{h}_k^v $, by correlating $r\left[ n \right]$ with $\widetilde{Gav}_{256}$ and $\widetilde{Gbv}_{256}$. 
 To reduce the impact of noise, we average $\widehat{h}_k^u $ and $\widehat{h}_k^v $ to obtain the final estimate 
 for the $k$-th CIR component as $\widehat{h}_k={{{\left( {\widehat{h}_k^u  + \widehat{h}_k^v } \right)}
 \mathord{\left/ {\vphantom {{\left( {\widehat{h}_k^u  + \widehat{h}_k^v } \right)} 2}} \right.
 \kern-\nulldelimiterspace} 2}}$. 

For ease of presentation, we assume the signal is transmitted with unit power in the above discussion. The effect of 
actual transmit power $P_t$ can be easily incorporated by scaling $h_k$ with a factor of $\sqrt {P_t}$. In practice, the quality 
of the estimated CIR component $\widehat{h}_k $ is affected by noise and other non-resolvable CIR components 
with respect to the $k$-th CIR component, which could, in turn, affect the performance of \name. As mentioned in 
\cite{sur201560}, the channel at 60GHz is inherently sparse with a few dominant paths. Thanks to the high 
transmission bandwidth, two CIR components can be resolved as long as the difference in their time of arrival is larger 
than $0.57ns$. Thus, the impact of non-resovlable CIR components on $\widehat{h}_k $ is expected to be limited if 
it corresponds to a dominant path. In Section IV, we will discuss how noise affects the performance of \nameS 
and show that non-resolvable CIR components have only limited effect on \name.


\section{Design of \name}
\nameS addresses the problem of estimating the AOA/AOD of a specific dominant path, given a set of beam-specific 
CIR measurements. How to identify the dominant paths of a channel is out of the scope of \name. As mentioned in 
\cite{ghasempour2018multi,pefkianakis2018accurate}, the dominant paths of a channel can be identified by detecting 
and aggregating the peaks in the given set of beam-specific CIR measurements. Without loss of generality, we assume 
$\rm K$ dominant paths, indexed as $\{1, \cdots, {\rm K}\}$, are identified from the given set of beam-specific CIR 
measurements, and we are interested in the angular information of the $\kappa$-th ($\kappa \in \{1, \cdots, {\rm K}\}$) 
dominant path. For simplicity, we call the CIR component corresponding to the $\kappa$-th dominant path as the 
$\kappa$-th dominant CIR component.

As mentioned before, \nameS exploits the CIR measured under different beams to infer the AOAs/AODs of dominant 
paths. These CIR measurements can be collected using, for example, sector sweep (SSW) frames during the 
sector-level sweep phase or a specific beam-sweeping process 
\cite{pefkianakis2018accurate,ghasempour2018multi,zhou2017beam}. In general, the beam-sweeping is
carried out in one or multiple stages. During each stage, one party fixes its beam pattern and the other party sweeps 
through different beams. For ease of presentation, we focus on AOA estimation in the following development and thus 
are interested in the case where the transmitter fixes its beam and the receiver receives the beam training frames, 
such as the SSW frames, with different beams. By following the procedure in \cite{pefkianakis2018accurate},
\nameS presented in this section can also be applied in AOD inference using the transmitter-side beam sweeping. 
Suppose the receiver can sweep through a maximum of $L$ beams, and thus we have the CIRs measured under 
$L$ different beams. Let $h_\kappa\left( l \right)$ be the value of the $\kappa$-th dominant CIR component when the 
$l$-th beam is used for reception. As shown in \cite{ghasempour2018multi,pefkianakis2018accurate}, the estimate of 
$h_\kappa\left( l \right)$, denoted as $\widehat{h}_\kappa\left( l \right)$, can be extracted from the estimated CIR 
under the $l$-th receiving beam based on the time of arrival of the $\kappa$-th dominant path.

Let  $\theta_\kappa$ be the AOA of the $\kappa$-th dominant path. To infer $\theta_\kappa$ from 
$\widehat{h}_\kappa\left( l \right)$'s, we need to draw a quantity from $\widehat{h}_\kappa\left( l \right)$'s so that 
it is determined solely by $\theta_\kappa$. Clearly from Eq.~(\ref{1}), it is difficult to directly infer $\theta _\kappa$ 
from the amplitudes of $h_\kappa\left( l \right)$'s since they are also affected by many other factors, such as the path 
gain and the directional gain of the transmit beam. Notice that, when the transmit power and transmit beam pattern are fixed, 
$h_\kappa\left( l \right)$'s satisfy:
\begin{align}
\label{13}
{{\left| {h_\kappa \left( {l_1 } \right)} \right|} \mathord{\left/
 {\vphantom {{\left| {h_\kappa \left( {l_1 } \right)} \right|} {\left| {h_\kappa \left( {l_2 } \right)} \right|}}} \right.
 \kern-\nulldelimiterspace} {\left| {h_\kappa \left( {l_2 } \right)} \right|}} = {{\left| {G_{l_1 } \left( {\theta _\kappa } \right)} \right|} \mathord{\left/
 {\vphantom {{\left| {G_{l_1 } \left( {\theta _\kappa } \right)} \right|} {\left| {G_{l_2 } \left( {\theta _\kappa } \right)} \right|}}} \right.
 \kern-\nulldelimiterspace} {\left| {G_{l_2 } \left( {\theta _\kappa } \right)} \right|}}.
\end{align}
In other words, when the $l_1$-th and the $l_2$-th beams are used for reception, the ratio of ${\left| {h_\kappa 
\left( {l_1 } \right)} \right|}$ to ${\left| {h_\kappa \left( {l_2 } \right)} \right|}$ is completely determined by 
$\theta _\kappa$. This observation leads us to use the ratio between $\left|\widehat{h}_\kappa\left( l \right)\right|$'s, 
instead of their absolute values, for AOA estimation.

Due to the very irregular beam pattern of 802.11ad devices, it is challenging to analytically derive the inverse 
function of ${{\left| {G_{l_1 } \left( {\theta _\kappa } \right)} \right|} \mathord{\left/
 {\vphantom {{\left| {G_{l_1 } \left( {\theta _\kappa } \right)} \right|} {\left| {G_{l_2 } \left( {\theta _\kappa } \right)} \right|}}} 
 \right.
 \kern-\nulldelimiterspace} {\left| {G_{l_2 } \left( {\theta _\kappa } \right)} \right|}}$ for AOA estimation. In view of Eq.~(\ref{13}), we address this challenge by searching for the direction which minimizes the difference between the 
 left-hand and right-hand sides of Eq.~(\ref{13}). Specifically, given ${\widehat{h}_\kappa  \left( {l_1 } \right)}$ and 
 ${\widehat{h}_\kappa  \left( {l_2 } \right)}$, we have
\begin{align}
\label{14}
\widehat\theta _\kappa   = \mathop {\arg \min }\limits_{\theta \in \Theta }  \left| {\frac{{\left| {\widehat{h}_\kappa  \left( {l_1 } \right)} \right|}}{{\left| {\widehat{h}_\kappa  \left( {l_2 } \right)} \right|}} - \frac{{\left| {G_{l_1 } \left( {\theta  } \right)} \right|}}{{\left| {G_{l_2 } \left( {\theta  } \right)} \right|}}} \right|,
\end{align}
where ${\widehat\theta _\kappa  }$ is the estimated AOA of the $\kappa$th dominant path, and $\Theta$ represents 
the search space. When new CIR measurements are available, we can combine them with ${\widehat{h}_\kappa  
\left( {l_1 } \right)}$ and ${\widehat{h}_\kappa  \left( {l_2 } \right)}$ to refine our estimation, which leads to \name.

To infer the AOA of the $\kappa$-th path, \nameS maintains a score for each direction. \nameS does not assume the 
order in which ${\widehat{h}_\kappa  \left( {1 } \right)},\cdots,{\widehat{h}_\kappa  \left( {L } \right)}$ are used for 
updating the score, but, in this section, we assume they are sequentially used for score updating for ease of 
presentation. Let $s_\kappa^{\ell}  \left( \theta  \right)$ be the score assigned to each direction based on 
${\widehat{h}_\kappa  \left( {1 } \right)},\cdots,{\widehat{h}_\kappa  \left( {\ell } \right)}$. Observing that the score is 
updated based on the variation between different CIR measurements, we have $s_\kappa^{1}  \left( \theta  \right)=0$, 
$\forall \theta \in \Theta$. For $\ell \ge 2$, $s_\kappa^\ell  \left( \theta  \right)$ can be obtained from 
$s_\kappa^{\ell-1}  \left( \theta  \right)$ using ${\widehat{h}_\kappa  \left( {\ell } \right)}$ as
\begin{align}
\label{15}
&s_\kappa^\ell  \left( \theta  \right) \nonumber \\=& \left\{ {\begin{array}{*{20}c}
   {s_\kappa^{\ell-1}  \left( \theta  \right) + \sum\limits_{l \in \mathcal{L}} {\left\{ {\begin{array}{*{20}c}
   {\alpha \left( {\ell ,l,\kappa ,\theta } \right) \times }  \\
   {\left| {\frac{{\left| {\widehat{h}_\kappa  \left( \ell  \right)} \right|}}{{\left| {\widehat{h}_\kappa  \left( l \right)} \right|}} - \frac{{\left| {G_\ell  \left( {\theta _\kappa  } \right)} \right|}}{{\left| {G_l \left( {\theta _\kappa  } \right)} \right|}}} \right|}  \\
\end{array}} \right\}} } & {\mathcal{L} \ne \emptyset ,\ell  \notin \mathcal{L}}  \\
   {s_\kappa  \left( \theta  \right)} & {\rm{otherwise}}  \\
\end{array}} \right.,
\end{align}
where $\mathcal{L}$ is the set of receiving beams used in the previous CIR measurement, and $\emptyset$ is an 
empty set. Initially, we have $\mathcal{L}=\{1\}$. ${\alpha \left( {\ell ,l,\kappa ,\theta } \right)}$'s are weight factors 
which allows us to weight the contribution of every two CIR measurements based on the corresponding beams, the 
considered direction, and the dominant path of interest. In this paper, we set $\alpha \left( {\ell ,l,\kappa ,\theta } \right)$ 
as $1\left( {\left| {\widehat{h}_\kappa  \left( \ell  \right)} \right| \ge \hbar } \right)1\left( {\left| {\widehat{h}_\kappa  
\left( l \right)} \right| \ge \hbar } \right)$, where $1\left( . \right)$ is the indicator function. $\hbar$ is a threshold and 
can be set to, for example, $\nu \left|\widehat{h}\right|_\kappa ^{\max }$, where $\nu \in \left[0,1\right]$, and $\left|
\widehat{h}\right|_\kappa ^{\max }  = \mathop {\max }\limits_{l \in \left\{ {1, \cdots ,L} \right\}} \left|\widehat{h}_\kappa  
\left( l \right)\right|$. We adopt such weight factors since the value of ${\left| {\widehat{h}_\kappa  \left( l \right)} \right|}$ 
cannot be fully trusted when it is too small. On the one hand, the small ${\left| {\widehat{h}_\kappa  \left( l \right)} \right|}
$ could be caused by either a low beamforming gain along $\theta_\kappa$ or an abrupt blockage of the $\kappa$-th 
dominant path. On the other hand, a small ${\left| {\widehat{h}_\kappa  \left( l \right)} \right|}$ is more susceptible to 
noise and might not be able to accurately track the variations in $\left|h_\kappa  \left( l  \right)\right|$. In other words, 
when either ${\left| {\widehat{h}_\kappa  \left( \ell \right)} \right|}$ or ${\left| {\widehat{h}_\kappa  \left( l \right)} \right|}$ is 
small, the variation in the beamforming gain along $\theta_\kappa$ might not be the major reason for the variation in 
${{\left| {\hat h_\kappa  \left( \ell  \right)} \right|} \mathord{\left/
 {\vphantom {{\left| {\hat h_\kappa  \left( \ell  \right)} \right|} {\left| {\hat h_\kappa  \left( l \right)} \right|}}} \right.
\kern-\nulldelimiterspace} {\left| {\hat h_\kappa  \left( l \right)} \right|}}$, and applying these ${\left| {\widehat{h}_\kappa  
\left( l \right)} \right|}$'s in AOA estimation could adversely affect the accuracy. Besides limiting the impact of small 
${\left| {\widehat{h}_\kappa  \left( l \right)} \right|}$'s, ${\alpha \left( {\ell ,l,\kappa ,\theta } \right)}$ also allows \nameS 
to incorporate prior information. For example, if $\theta_\kappa$ is known to be within a direction range 
$\widetilde\Theta$, we can exploit such information for AOA estimation by assigning $1$ to 
${\alpha \left( {\ell ,l,\kappa ,\theta } \right)}$, $\forall \theta \in \widetilde \Theta$, and $0$ otherwise. 

Once the updated score $s_\kappa^\ell  \left( \theta  \right)$ is obtained, we update $\mathcal{L}$ as
\begin{align}
\label{16}
\mathcal{L} =\mathcal{L} \cup \left\{ \ell  \right\}.
\end{align}

After going through all $L$ CIR measurements, \nameS outputs the estimated AOA of the $\kappa$-th dominant 
path as
\begin{align}
\label{17}
\widehat\theta _\kappa   = \mathop {\arg \min }\limits_{\theta  \in \Theta } s_\kappa^L  \left( \theta  \right).
\end{align}

\nameS is summarized  in Algorithm 1. 
\begin{center}
\begin{algorithm}[!t]
\caption{: \name}
\begin{algorithmic}[1]
\REQUIRE ${\widehat{h}_\kappa  \left( l \right)}$, $l \in \{1,\cdots,L\}$, ${G_l \left( {\theta  } \right)}$, $l \in \{1,\cdots,L\}$, $\theta \in \Theta$, $\Theta$, $\hbar$
\ENSURE $\widehat\theta _\kappa$
\STATE $s_\kappa ^1 \left( \theta  \right) = 0$, $\forall \theta \in \Theta$, $\mathcal{L}=\{1\}$
\FOR{$\ell$=2 to $L$}
  \FOR{$l \in \mathcal{L}$}
  \STATE Determine ${\alpha \left( {\ell ,l,\kappa ,\theta } \right)}$ based on ${\widehat{h}_\kappa  \left( \ell \right)}$, ${\widehat{h}_\kappa  \left( l \right)}$, and other information (e.g., $\hbar$ in this paper)
  \ENDFOR
  \STATE $\forall \theta \in \Theta$, update $s_\kappa ^{\ell-1}\left( \theta  \right)$ to $s_\kappa ^\ell  
  \left( \theta  \right)$ according to Eq.~(\ref{15})
  \STATE $\mathcal{L} = \mathcal{L} \cup \left\{ \ell  \right\}$
\ENDFOR
\STATE $\widehat\theta _\kappa   = \mathop {\arg \min }\limits_{\theta  \in \Theta } s_\kappa ^L \left( \theta  \right)$
\end{algorithmic}
\end{algorithm}
\end{center}

Given $\widehat\theta _\kappa$, we are interested in whether it is an accurate estimate of $\theta_\kappa$. 
This information is helpful since it enables us to determine, for example, whether we should collect more 
beam-specific measurements to refine $\widehat\theta _\kappa$. Without noise and non-resolvable multipath 
components, $\theta_\kappa$ should satisfies Eq.~(\ref{13}). Considering the potential impacts of noise and 
non-resolvable multipath components, we propose to determine if $\widehat\theta _\kappa$ is an accurate 
estimate of $\theta _\kappa$ through the following inequality:
\begin{align}
\label{18}
 \left| {\frac{{\left| {\widehat{h}_\kappa  \left( {\underline l  } \right)} \right|}}{{\left| {\widehat{h}_\kappa  \left( {\overline l  } \right)} \right|}} - \frac{{\left| {G_{\underline l } \left( {\theta  } \right)} \right|}}{{\left| {G_{\overline l } \left( {\theta  } \right)} \right|}}} \right| \le \varepsilon,
\end{align}
where $\overline l \mathop { = \arg \max }\limits_{l \in \left\{ {L - 1,L} \right\}} \widehat{h}_\kappa  \left( l \right)$, $
\underline l \mathop { = \arg \min }\limits_{l \in \left\{ {L - 1,L} \right\}} \widehat{h}_\kappa  \left( l \right)$, and $
\varepsilon  \in \left[ {0,1} \right]$ is a threshold. We use $\overline l$ and $\underline l$, instead of $L-1$ and $L$, in 
Eq.~(\ref{18}) so that we can employ the same $\varepsilon$ no matter $\left|\widehat{h}_\kappa  \left( {L - 1} \right)
\right| \ge \left|\widehat{h}_\kappa  \left( L \right)\right|$ or $\left|\widehat{h}_\kappa  \left( {L} \right)\right| \ge \left|
\widehat{h}_\kappa  \left( L-1 \right)\right|$. The effectiveness of this criterion will be evaluated in the next section 
using simulations.
 
\section{Performance Evaluation}
\subsection{Simulation Setup and Evaluation Methodology}

We now evaluate the performance of \nameS based on beam patterns measured on the Dell D5000 docking station 
and the TP-Link Talon AD7200 router, which are off-the-shelf 60GHz devices with phased array antenna and beam 
sweeping implementation. The $32$-beam patterns swept by D5000 docking stations for device discovery and the $35$ 
default beams used by Talon routers are provided in \cite{bielsa201860} and \cite{steinmetzer2017compressive}. 
In this evaluation, we process the measurement data in \cite{bielsa201860} and \cite{steinmetzer2017compressive} 
so that the highest directional power gain among all beams of each device is $15dB$. Our evaluation is conducted 
in a $4\times 3$m room. The signal propagation in this environment follows the model suggested by the IEEE Task 
Group ad, which is validated by experimental results \cite{IEEE2010Channel}. In this model, signal paths are generated 
by exploiting the clustering phenomenon in a 60GHz channel. Specifically, the first- and second-order reflections are 
first generated using ray-tracing techniques \cite{steinmetzer2016mmtrace}, and these paths are then blurred to 
generate path clusters. For each path identified via ray-tracing, a path cluster is generated by following the procedure
(Section 3.7) and the parameter settings (Table 10) presented in \cite{IEEE2010Channel}. Besides multipath 
propagation, the received signal is also affected by an additive Gaussian noise. We assume the CIRs are extracted 
from the SSW frames, and thus the preamble is generated according to the control PHY. Following  
\cite{IEEE2012adIEEE}, the signal is transmitted over a band with center frequency $60.48GHz$ and bandwidth 
$1760MHz$. We are interested in the AOAs of dominant paths and thus consider a receive sector sweep where the 
transmitter transmits SSW frames using the same beam pattern while the receiver sweeps over a set of beams to 
receive these SSW frames. The STF and CE field of these SSW frames are generated through Golay sequences 
and modulated using ${\pi  \mathord{\left/
 {\vphantom {\pi  2}} \right.
 \kern-\nulldelimiterspace} 2}$-BPSK according to the 802.11ad standard. Once the signal is received at the receiver, 
 we implement the channel estimation presented in Section II to extract the beam-specific CIR measurements, which 
 will be used in \nameS for AOA inference. It should be noted that the antenna arrays of the transmitter and 
 the receiver are facing each other in our simulation. 

\begin{figure*}[!t]
 \begin{center}
 \subfigure[Scenario A.]{
\includegraphics[width=0.23\textwidth]{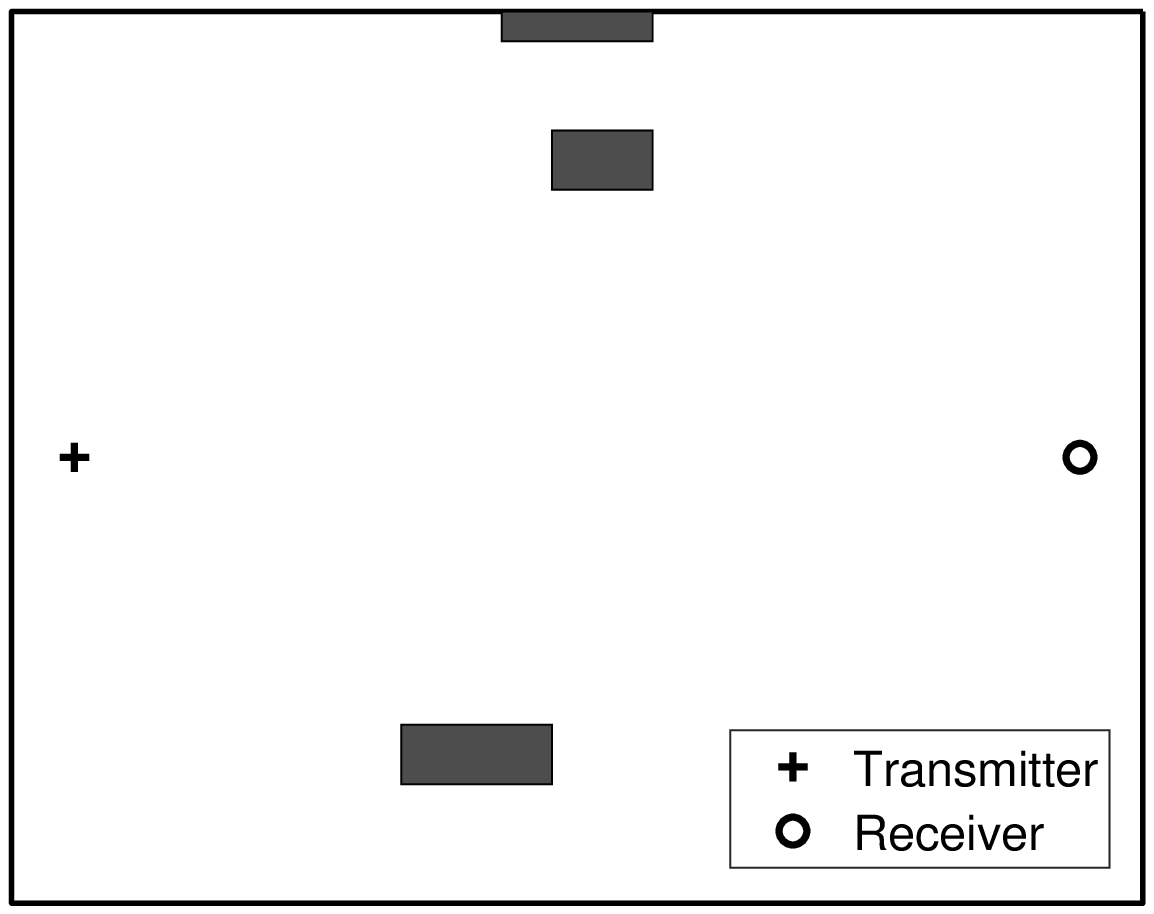}
}
\subfigure[Scenario B.]{
\includegraphics[width=0.23\textwidth]{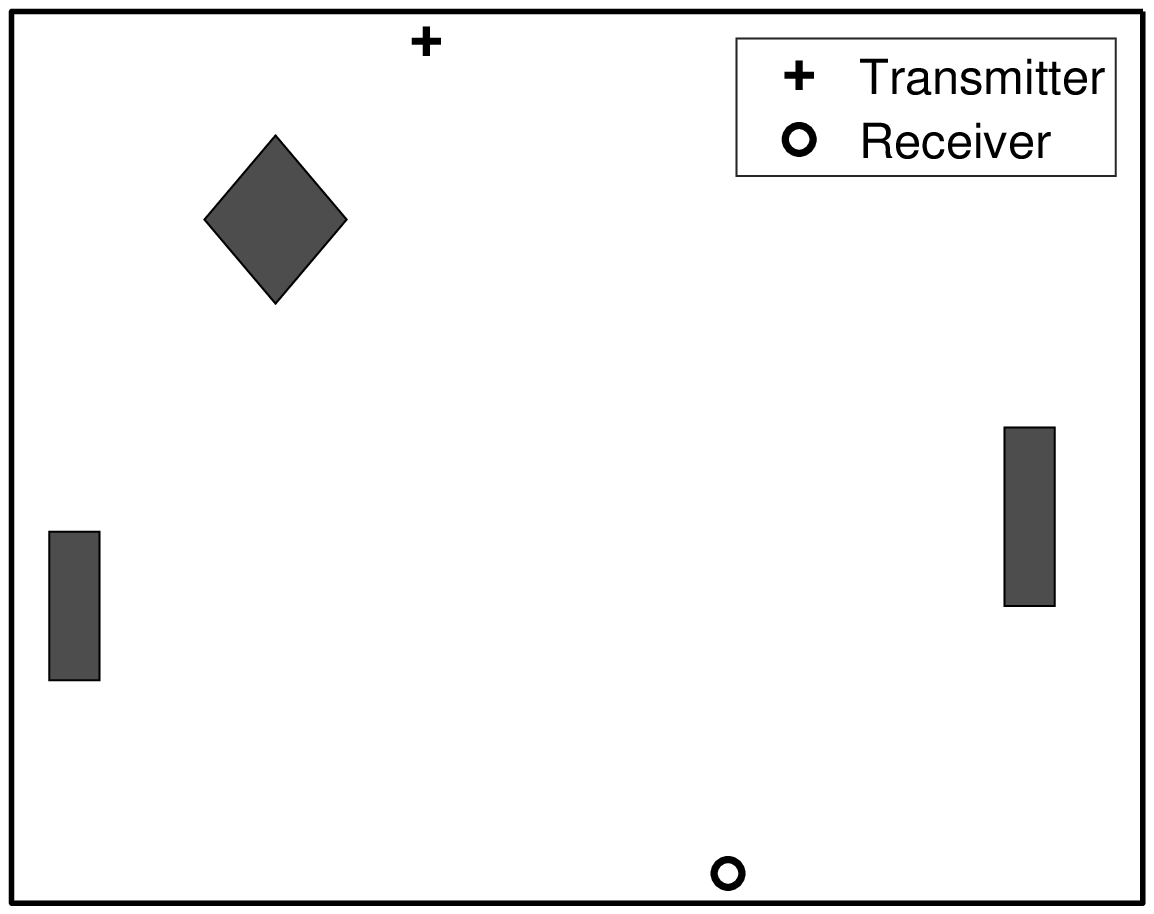}
}
\subfigure[CIR of Scenario A.]{
\includegraphics[width=0.23\textwidth]{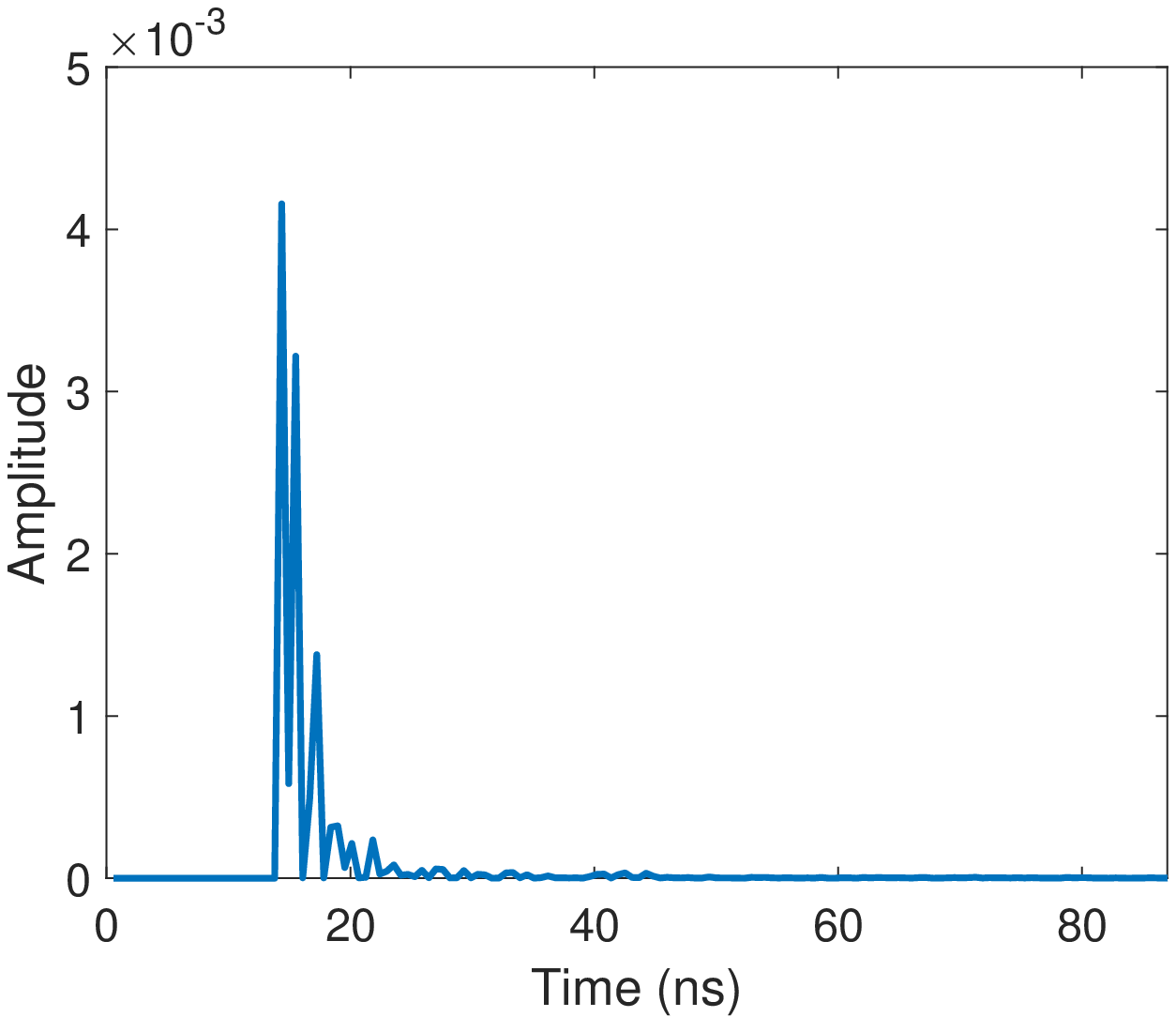}
}
\subfigure[CIR of Scenario B.]{
\includegraphics[width=0.23\textwidth]{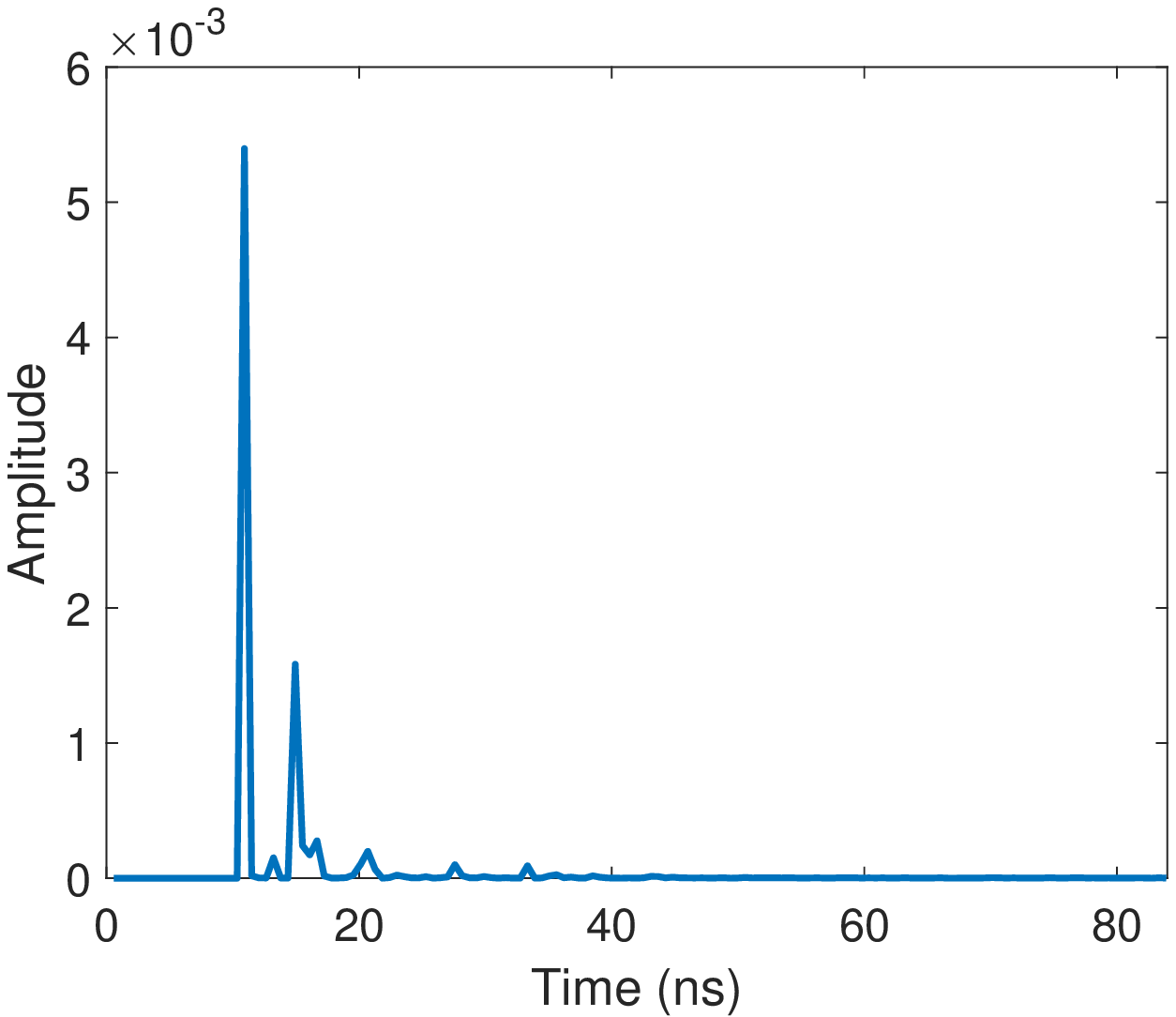}
}
  \end{center}
  \begin{center}
   \parbox{16cm}{\caption{The considered scenarios and the CIR measured in these scenarios. Black blocks in the 
   figure represent objects which could block and reflect the mmWave signals. (a) The transmitter locates at $\left[-2,0\right]$. The receiver locates at $\left[2,0\right]$. The three objects are characterized by $\left(0.1, 1, 0.4, 0.2, 0^\circ , 3.24\right)$, $\left(-0.4, -1, 0.6, 0.2, 180^ \circ, 3.24\right)$, $\left(0, 1.45, 0.6, 0.1, 0^ \circ, 3.24\right)$. (b) The transmitter locates at $\left[-0.6,1.4\right]$. The receiver locates at $\left[0.6,-1.4\right]$. The three objects are characterized by $\left(1.8,  -0.2, 0.2, 0.6, 0^\circ , 3.24\right)$, $\left(-2, -0.5, 0.2, 0.5, 180^ \circ, 3.24\right)$, $\left(-1.2, 0.8, 0.4, 0.4, 45^ \circ, 3.24\right)$. (c-d) The CIR measured in scenario A and B.}  \label{fig3}}
  \end{center}
\end{figure*}

In what follows, we will evaluate the performance of \nameS based on the two scenarios shown in 
Figs.~\ref{fig3}(a) and (b) where the transmitter and the receiver are characterized by their locations. 
We consider a coordinate system with the origin at the center of the room. The objects in the environment are 
characterized by six-element tuples where the first two elements record their locations, the next two elements 
present their length and width, the fifth element is their orientation, and the last one is their dielectric constant. 
The examples of beam-specific CIRs extracted from these two scenarios are shown in Figs.~\ref{fig3}(c) and 
(d) where the receiver uses the $2$nd beam measured on the D5000 docking station for reception and the 
transmitter adopts a quasi-omni pattern with a transmit power of $25dBm$. The dielectric constant of the wall is set to 
$2$ \cite{de2014remote}. Form Figs.~\ref{fig3}(c) and (d), we can observe $3$ peaks in Scenario A and $2$ peaks 
in Scenario B, which correspond to dominant paths in the environment. We will take the paths/peaks identified 
in Figs.~\ref{fig3}(c) and (d) as examples to evaluate the performance of \name.

\subsection{Result and Analysis}
 \begin{figure*}[!t]
 \begin{center}
 \subfigure[Scenario A, D5000, $1$st path.]{
\includegraphics[width=0.3\textwidth]{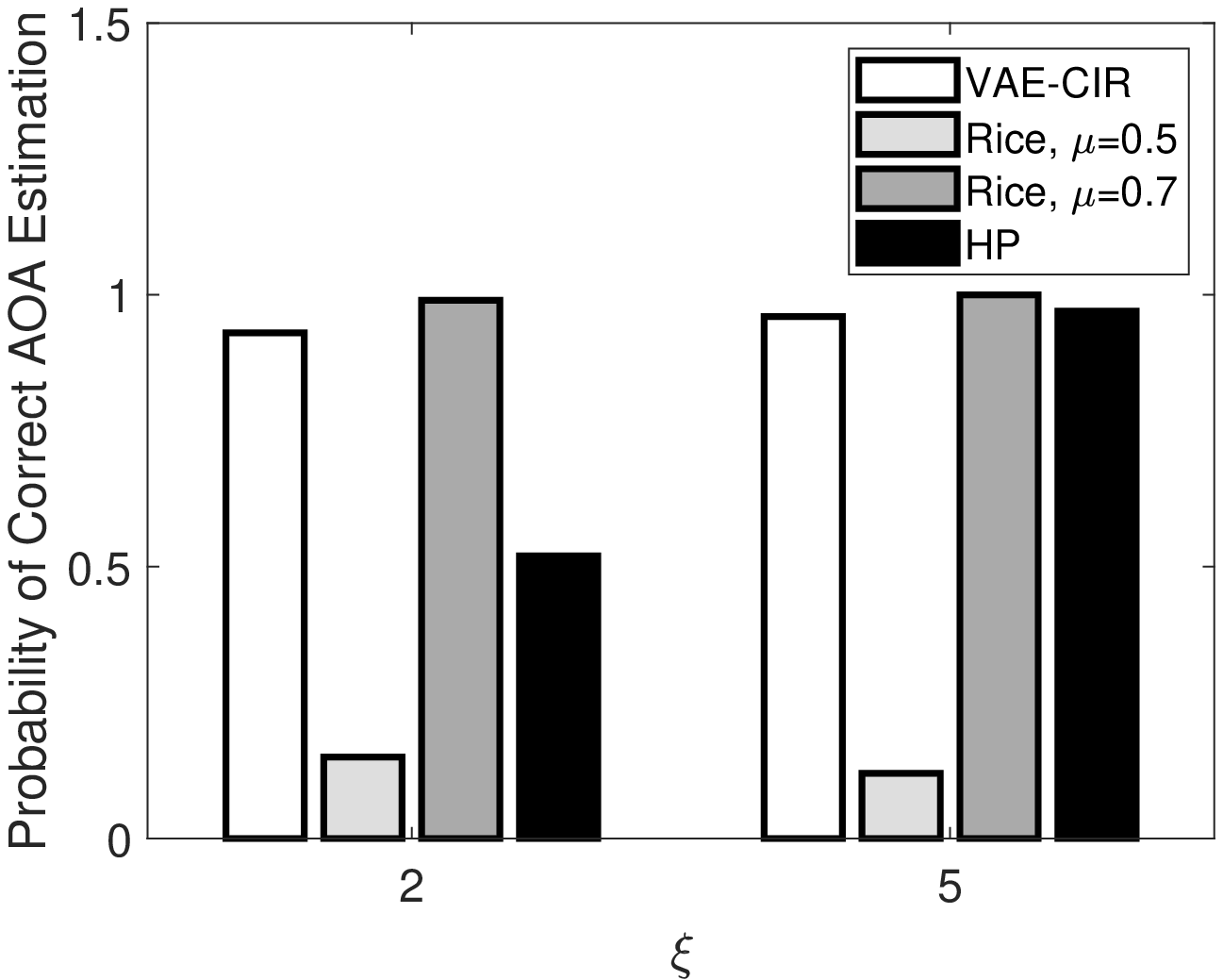}
}
\subfigure[Scenario B, D5000, $1$st path.]{
\includegraphics[width=0.3\textwidth]{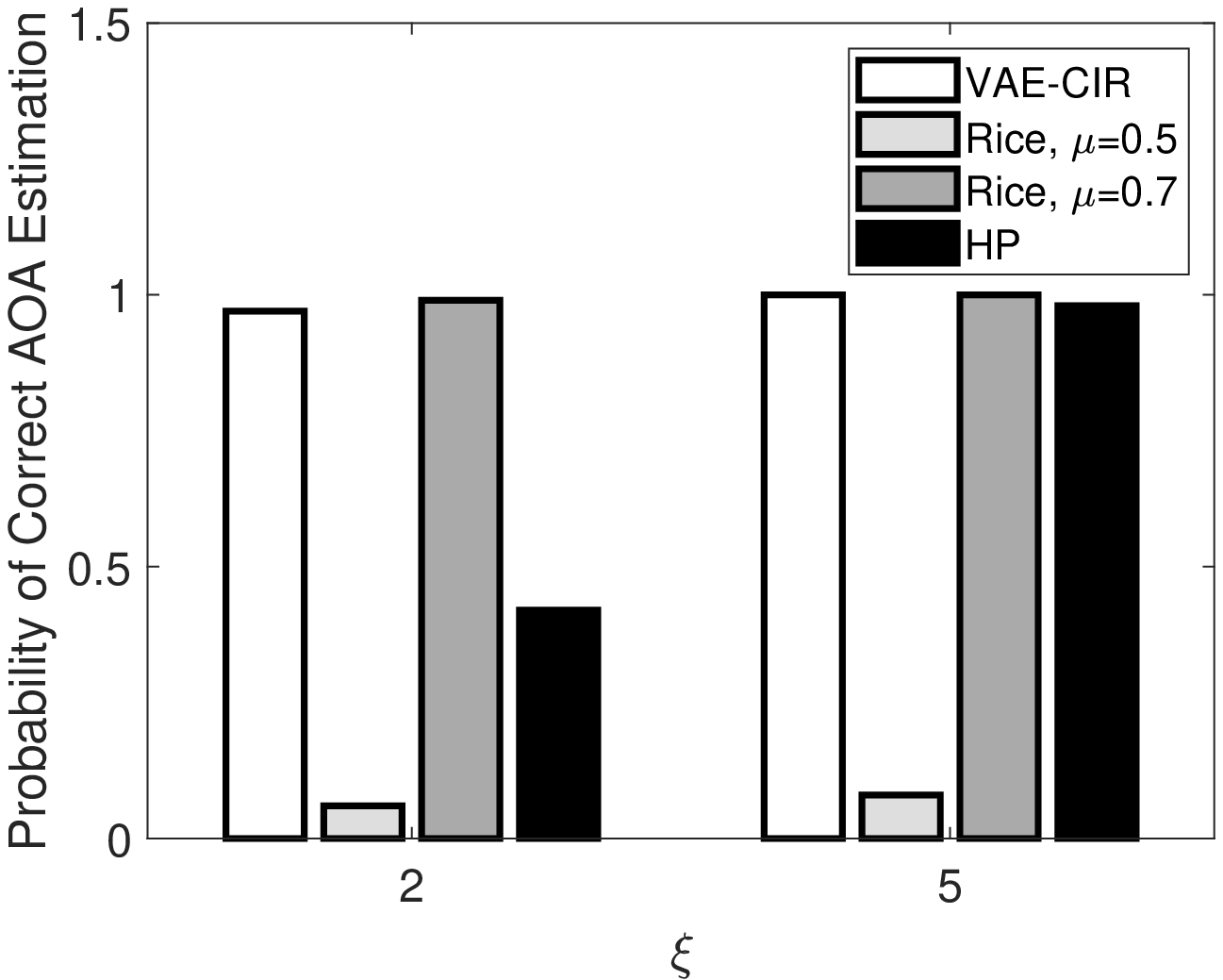}
}
\subfigure[Scenario A, D5000, $3$rd path.]{
\includegraphics[width=0.3\textwidth]{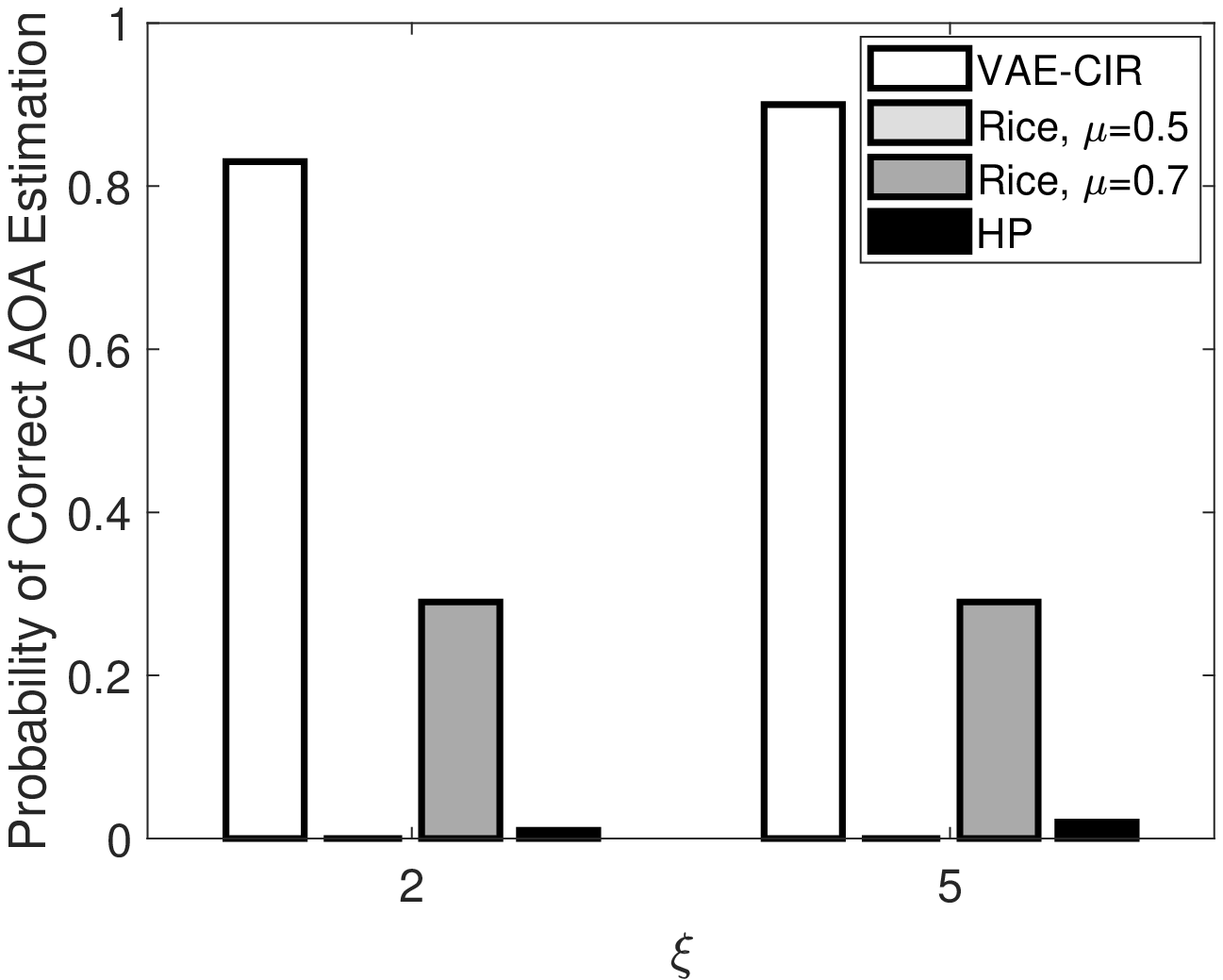}
}
\subfigure[Scenario B, D5000, $2$nd path.]{
\includegraphics[width=0.3\textwidth]{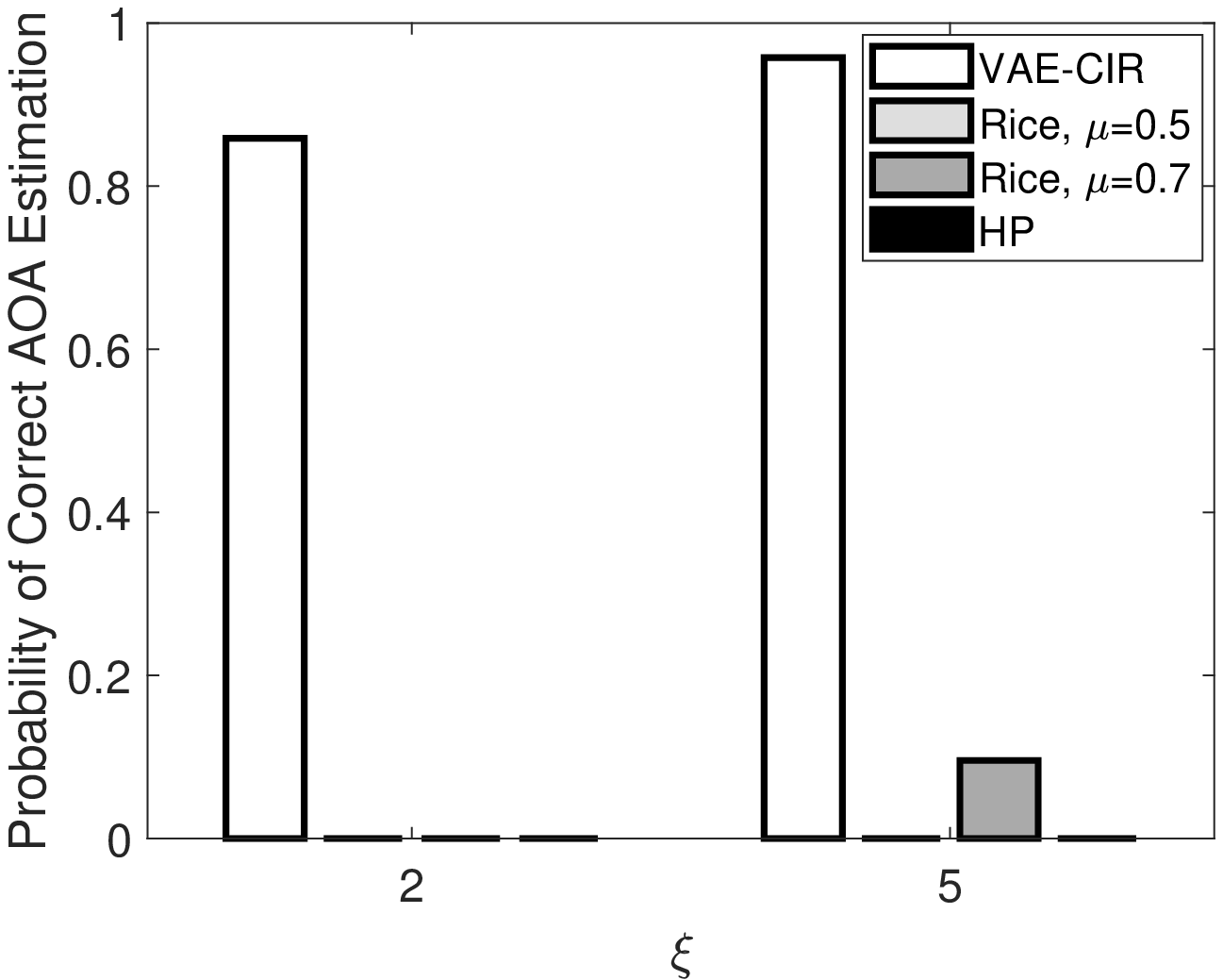}
}
\subfigure[Scenario A, Talon, $1$st path.]{
\includegraphics[width=0.3\textwidth]{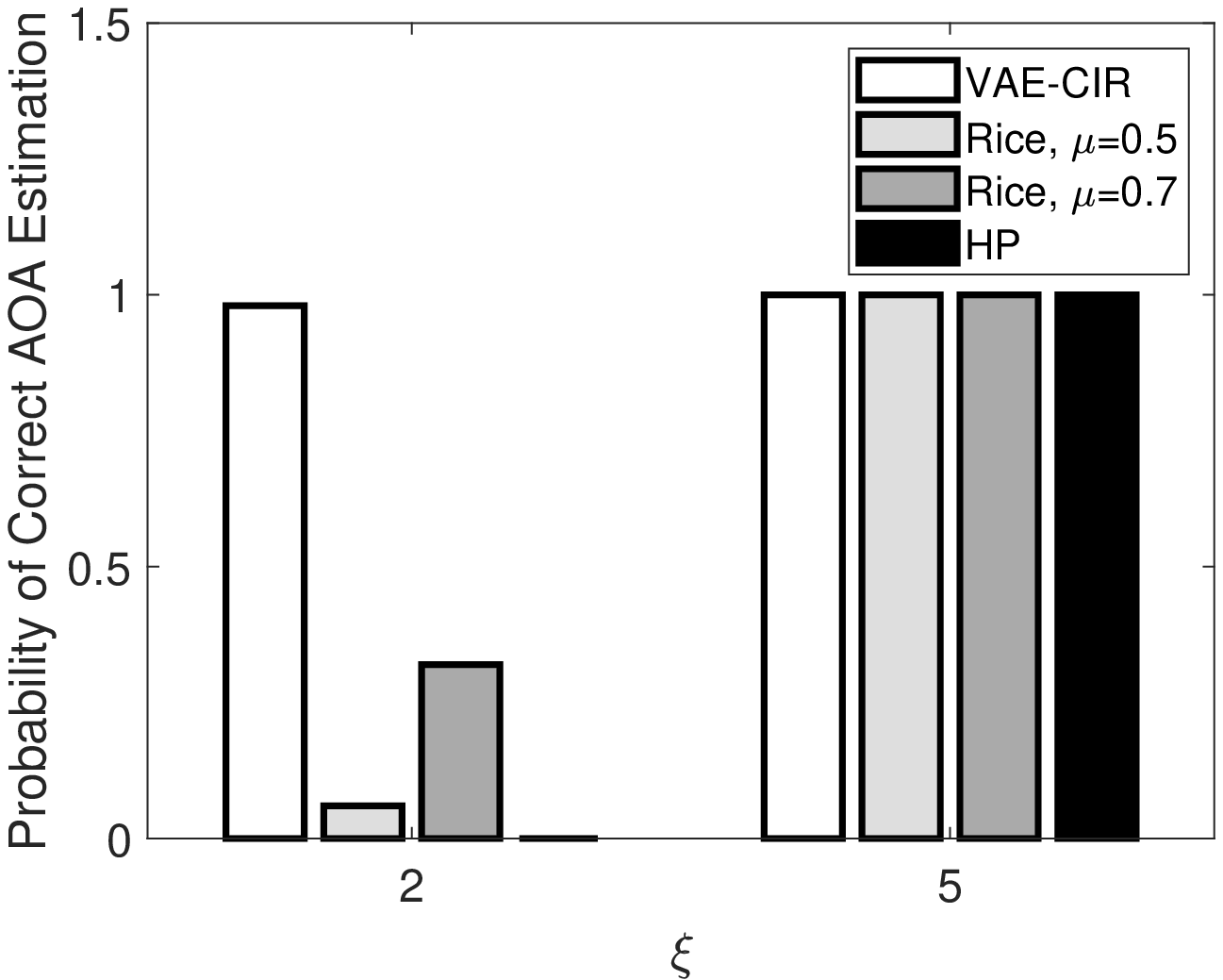}
}
\subfigure[Scenario B, Talon, $1$st path.]{
\includegraphics[width=0.3\textwidth]{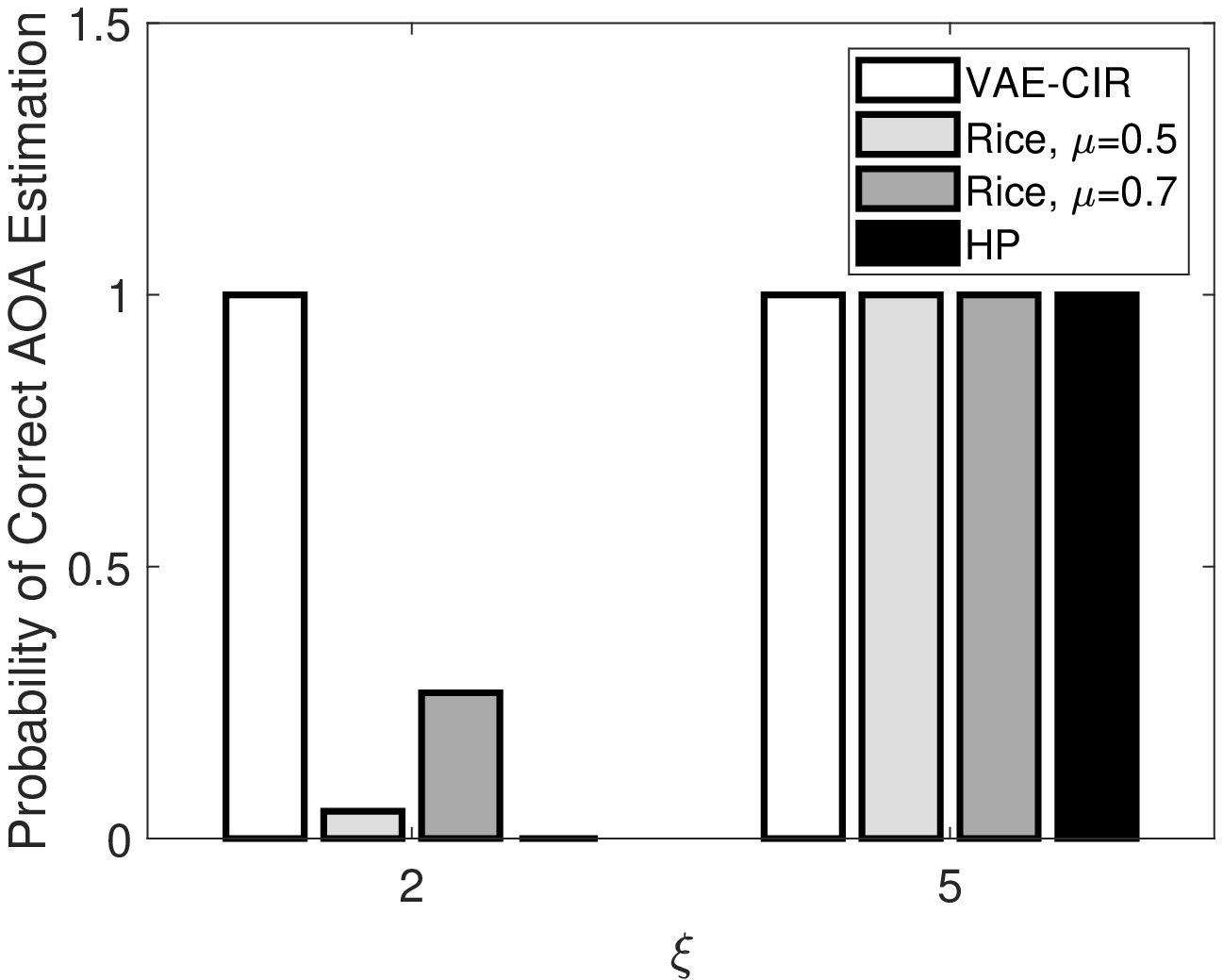}
}
  \end{center}
  \begin{center}
\parbox{10cm}{\caption{The probability of correct AOA estimation in different scenarios.}  \label{fig4}}
  \end{center}
\end{figure*}

Fig.~\ref{fig4} compares the probability of correct AOA estimation of \nameS with those of existing schemes. For effective comparison, we consider the angular inference schemes in \cite{ghasempour2018multi} and \cite{pefkianakis2018accurate}, which are referred to as ``Rice'' and ``HP'' in Fig. \ref{fig4}. Similar to \name, both of these schemes assign scores to different directions by correlating beam-specific CIR measurements with the corresponding beams in use and infer angular information based on the final scores of different directions. However, unlike \name, they exploit absolute beam-specific CIR measurements, instead of their variations under different beams, for score updating. Besides, these two schemes always assign larger score increments to the directions with higher directional power gain, instead of the directions offering better match between the variations in CIRs and the variations in directional gains. For \name, the threshold $\hbar$ in $\alpha \left( {\ell ,l,\kappa ,\theta } \right)$ is set to $\nu \left|\widehat{h}\right|_\kappa ^{\max }$, where $\nu=0.1$ and $\left|\widehat{h}\right|_\kappa ^{\max }  = \mathop {\max }\limits_{l \in \left\{ {1, \cdots ,L} \right\}} \left|\widehat{h}_\kappa  \left( l \right)\right|$. In other words, $\left|\widehat{h}_\kappa  \left( l \right)\right|$ will be used to infer $\theta_\kappa$ only if $\left|\widehat{h}_\kappa  \left( l \right)\right| \ge 0.1\left|\widehat{h}\right|_\kappa ^{\max } $. For the ``Rice'' scheme, the $\kappa$-th dominant CIR component is considered to be detected under the $l$-th beam if $\left|\widehat{h}_\kappa  \left( l \right)\right| \ge \mu \left|\widehat{h}\right|_\kappa ^{\max } $ \cite{ghasempour2018multi}. For the ``HP'' scheme, the five $\left|\widehat{h}_\kappa  \left( l \right)\right|$'s with the highest amplitudes are used for AOA inference and the estimated AOA is the center of the direction interval spanned by the five directions with highest final scores \cite{pefkianakis2018accurate}. In Fig. \ref{fig4}, the estimated AOA $\widehat\theta _\kappa$ is considered to be accurate as long as $\left| {\widehat\theta _\kappa   - \theta _\kappa  } \right| \le \xi $, where ${\theta _\kappa  }$ is the true AOA of the $\kappa$-th path. In our experiment, the signal is transmitted at a power of $25dBm$ using the quasi-omni pattern and propagates through a multipath channel illustrated in Section IV.A. We evaluate the performance of \nameS by using the beam patterns measured on different off-the-shelf 802.11ad devices (Dell D5000 docking station and TP-Link Talon AD7200 router) as the receiving beam patterns swept by the receiver\footnote{By viewing this receiver-side beam sweeping as the transmitter-side sweeping, our experiment can also be regarded as an evaluation of the AOD estimation performance of \nameS \cite{pefkianakis2018accurate}.}. Our following evaluations are based on those paths identified in Fig. \ref{fig3}. First, we investigate the performance of \nameS in Scenario A and B with the beam patterns measured from the Dell D5000 docking station. In Fig. \ref{fig4}(a) and \ref{fig4}(b), we present the AOA estimation results for the first dominant paths in Scenario A and B with $-30dBm$ noise in the channel. In Fig. \ref{fig4}(c) and \ref{fig4}(d), we are interested in the AOA of the third path in Scenario A and the AOA of the second path in Scenario B, where the channel noise is set to $-40dBm$ and $-45dBm$, respectively. Then, we redo the experiment for the first paths in Scenario A and B, but, using the beam patterns measured from the Talon router. The obtained results are shown in Fig. \ref{fig4}(e) and \ref{fig4}(f) where the noise power is set to $-40dBm$. From Fig. \ref{fig4}, although the ``Rice'' and ``HP'' schemes achieve reasonable performance in a few cases, they do not performance well in other cases. In contrast, \nameS can achieve good AOA estimation results in all these cases, which demonstrates its effectiveness and the limited impact of nonresovable CIR components on \name. As shown in Fig. \ref{fig4}, the accuracy of the ``Rice'' scheme highly depends on the value of $\mu$, and both the ``Rice'' and ``HP'' schemes do not perform well for the $3$rd path in Scenario A and the $2$nd path in Scenario B. All of these are resulted from their score updating mechanism. Based on the values of $\left|\widehat{h}_\kappa  \left( l \right)\right|$'s, these schemes determine whether the corresponding beam patterns should be used for score increment. For each selected beam pattern, they assign higher score increment to the direction with higher directional power gain and thus are likely to output an accurate AOA estimation if the beams used for score updating have strong side lobes around the angle of interest. Clearly, if $\left|\widehat{h}_\kappa  \left( l \right)\right|$ is small, the $l$-th beam is unlikely to have strong side lobes along $\theta_\kappa$. In this case, exploiting $\left|\widehat{h}_\kappa  \left( l \right)\right|$ and the $l$-th beam for score increment would lead to erroneous score increase along the direction around the strong side lobes of the $l$-th beam. This explains why the performance of the ``Rice'' scheme degrades with a smaller $\mu$. By examining the beam patterns of the D5000 docking station, we notice that most of its beams do not have strong side lobes around the AOAs of the $3$rd path in Scenario A and the $2$nd path in Scenario B, and thus the ``Rice'' and the ``HP'' schemes do not assign high enough score increments to the directions of interest, which eventually leads to inaccurate AOA estimation. In other words, since the ``Rice'' and the ``HP'' schemes select beam patterns for score increments based on the absolute values of $\left|\widehat{h}_\kappa  \left( l \right)\right|$'s, which are not closely related to the direction of interest, their accuracy highly depends on the match between selected beam patterns and path directions. This observation further validates \nameS which uses the variations in $\left|\widehat{h}_\kappa  \left( l \right)\right|$'s, the quantities closely related to the direction of interest, for AOA estimation.

\begin{figure}[!t]
 \begin{center}
 \subfigure[Scenario A.]{
\includegraphics[width=0.35\textwidth]{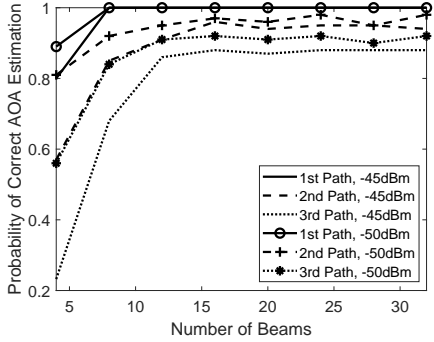}
}
\subfigure[Scenario B.]{
\includegraphics[width=0.35\textwidth]{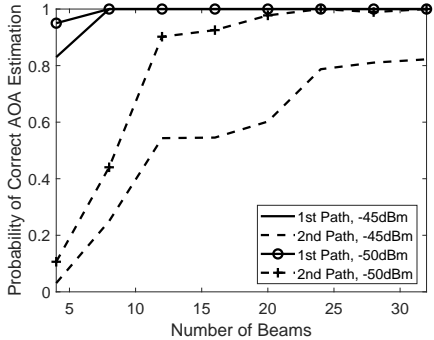}
}
  \end{center}
  \begin{center}
\parbox{8cm}{\caption{The probability of correct AOA estimation v.s. additive noise and the number of beams used for CIR measurement.}  \label{fig5}}
  \end{center}
\end{figure}

\begin{figure}[!t]
 \begin{center}
\includegraphics[width=0.35\textwidth]{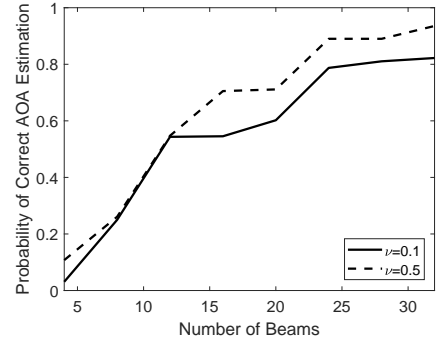}
  \end{center}
  \begin{center}
\parbox{8cm}{\caption{The probability of correct AOA estimation v.s. weight factors.}  \label{fig6}}
  \end{center}
\end{figure}

Then, we investigate how the performance of \nameS is affected by the additive noise and the number of beams used for measurement. Similar to Fig. \ref{fig4}, we assume the signal is transmitted at $25dBm$ using the quasi-omni pattern, and $\widehat\theta _\kappa$ is considered to be accurate as long as $\left| {\widehat\theta _\kappa   - \theta _\kappa  } \right| \le 2$. The noise power is set to the value shown in the legend of Fig. \ref{fig5}. The results shown in Fig. \ref{fig5} are obtained based on the $32$ beams of the Dell D5000 docking station. From Fig. \ref{fig5}, the accuracy of AOA estimation generally improves when more beams are available for beam-specific CIR collection. With more beam-specific measurements, we could make more informative decisions and thus improve the accuracy of the AOA estimation. It can be observed from Fig. \ref{fig5} that more accurate angular inference can be achieved when there is less noise in the channel. Clearly from Section III, the accuracy of \nameS depends on how accurately $\left|\widehat{h}_\kappa  \left( l \right)\right|$ approximates $\left|h_\kappa  \left( l \right)\right|$. The derivation in Section II.B shows that $\left|\widehat{h}_\kappa  \left( l \right)\right|$ more accurately approximates $\left|h_\kappa  \left( l \right)\right|$ when there is less noise in the channel. This explains why the performance of \nameS improves with noise power decreasing. The impact of noise also can be shown through the observation that \nameS achieves different accuracy for different paths. For example, for the two paths identified in Fig. \ref{fig3}(d), we can achieve a higher probability of correct AOA estimation for the $1$st path (shown in Fig. \ref{fig5}(b)). From \ref{fig3}(d), the dominant CIR component corresponding to the $1$st path is much stronger than that corresponding to the $2$nd path and thus is less susceptible to noise, which eventually leads to a higher probability of correct AOA estimation. This can be further corroborated by the results in Fig. \ref{fig6} where we change $\nu$ from $0.1$ to $0.5$ and $\widehat{h}_\kappa  \left( l \right)$ will be used in score updating only when $\left|\widehat{h}_\kappa  \left( l \right)\right| \ge 0.5 \left|\widehat{h}\right|_\kappa ^{\max }$. In so doing, weak $\left|\widehat{h}_\kappa  \left( l \right)\right|$'s significantly distorted by noise are more likely to be excluded from AOA estimation, which reduces accumulation of noise effect and improves the probability of correct AOA estimation as shown in Fig. \ref{fig6}, particularly when a large number of beam-specific CIRs are used for angular inference. These observations also demonstrate the importance of introducing the weight factors ${\alpha \left( {\ell ,l,\kappa ,\theta } \right)}$'s to \name. It should be noted that a large $\nu$ will not always lead to a better performance since it might limit the number of CIR measurements exploited for AOA estimation. As shown in Fig. \ref{fig5}, in this case, we might not be able to accurately estimate the AOAs due to limited information. In practice, the value of $\nu$ should be carefully selected. How to choose the optimal $\nu$ is out of the scope this paper and will be left for future work.

Finally, we evaluate if the criterion introduced at the end of Section III can correctly determine if $\widehat\theta _\kappa$ is an accurate estimate of $\theta _\kappa$. Specifically, we redo the simulation in Fig. \ref{fig5} with $\nu=0.1$ and the $32$ beams of the D5000 docking station. Similar to Fig. \ref{fig5}, $\widehat\theta _\kappa$ is considered to be an accurate estimate of $\theta _\kappa$ if $\left| {\widehat\theta _\kappa   - \theta _\kappa  } \right| \le 2$. Each time \nameS outputs an estimated AOA, $\widehat\theta _\kappa$, we check if Eq.~(\ref{18}) is valid, with $\varepsilon$ set to $0.3$ to account for the potential impact of noise. If the inequality is valid, the criterion determines $\widehat\theta _\kappa$ to be an accurate estimate of $\theta _\kappa$. The decisions being made is then compared to the ground truth. The results are shown in Table \ref{table} where $P_F$ is the percentage that the proposed criterion misclassifies $\widehat\theta _\kappa$ as an accurate estimate of $\theta _\kappa$ and $P_D$ is the percentage that the proposed criterion correctly identify $\widehat\theta _\kappa$ as an accurate estimate of $\theta _\kappa$. From Table \ref{table}, the criterion in Eq.~(\ref{18}) can accurately determine if $\widehat\theta _\kappa$ is an accurate estimate of $\theta _\kappa$. As demonstrated in Fig. \ref{fig3}(d), the $2$nd path in Scenario B is weaker than the $1$st path. Hence, for the two paths in Scenario B, the criterion in Eq.~(\ref{18}) is more susceptible to noise and thus less accurate when applied to the $2$nd path.
\begin{table}[!t]
\begin{center}
\caption{The performance of the criterion in Eq.~(\ref{18})}
\label{table}
\begin{tabular}{|l|c|c|c|c|}
\hline
\multirow{2}{*}{\backslashbox{Path}{Metric}} &\multicolumn{2}{|c|}{$-45$dBm} & \multicolumn{2}{|c|}{$-50$dBm} \\
\cline{2-5}
& $P_F$ & $P_D$ & $P_F$ & $P_D$ \\
\cline{1-5}
Scenario A, $1$st &$0\%$ & $100 \%$ & $0\%$ & $100 \%$ \\
\cline{1-5}
Scenario A, $2$nd & $0 \%$ & $99 \%$ & $0\%$ &$99\%$ \\
\cline{1-5}
Scenario A, $3$rd & $2\%$ & $100\%$ & $1\%$ & $100\%$ \\
\cline{1-5}
Scenario B, $1$st & $0\%$ & $100\%$ & $0 \%$ & $100\%$\\
\cline{1-5}
Scenario B, $2$nd & $8.43\%$ & $91.57\%$ & $1.06\%$ & $98.94\%$ \\
\hline
\end{tabular}
\end{center}
\end{table}

\section{Conclusion}
In this paper, we propose an angle estimation scheme, called \name, for 802.11ad devices by exploiting beam-specific CIR measurements. Unlike existing work, we exploit the variations between the CIRs measured under different beams, instead of their absolute values, for angle estimation. To evaluate the performance of \name, we simulate the beam sweeping operation of 802.11ad devices with beam patterns measured on off-the-shelf 802.11ad devices and the ray-tracing based channel generation. Our experimental results show that \nameS can enable accurate angular inference for 802.11ad devices and demonstrate its superiority to existing schemes. We expect \nameS to enable various applications, such as link performance prediction and device tracking, on 802.11ad devices when combined with other proposals for 60GHz networking.
\renewcommand\refname{References}
\bibliographystyle{IEEEtran}
\bibliography{beamselection}
\end{document}